# Pure Risk*

David Dillenberger[†] Jay Lu[‡]

August 19, 2026

**Abstract**

We introduce a behavioral notion of domain-specific risk aversion that separates attitudes toward risk from deterministic utility: an agent is more pure risk averse in one domain than in another if, for prizes that are indifferent under certainty, he is more averse to risk in the former domain than in the latter. We develop a model that goes beyond expected utility by allowing risk attitudes to vary across domains, while preserving expected utility within each domain. The domains are subjective and need not be specified in advance; they are identified from choice behavior. We establish uniqueness of the model's parameters and provide an axiomatic characterization.

*We thank Tommaso Denti, Mira Frick, Faruk Gul, Kevin He, George Mailath, Efe Ok, Luciano Pomatto, Rani Spiegler, and Mu Zhang for helpful comments.

[†]Department of Economics, University of Pennsylvania (email: ddill@sas.upenn.edu).

[‡]Department of Economics, University of California, Los Angeles (email: jay@econ.ucla.edu).

# 1 Introduction

Understanding risk attitudes is central to numerous economic analyses. Demand for insurance, portfolio choices, participation in extreme sports, and medical decision-making are only a few examples of activities directly shaped by individuals' willingness to bear risk. In the standard expected utility framework, each consequence is assigned a utility value, and a lottery is evaluated by the expected utility of its consequences. This implies that two outcomes with identical utility levels are interchangeable: replacing one with the other in any lottery does not affect the lottery's evaluation. For instance, if $u(x) = u(y)$, then, for any prize $z$, the decision maker should be indifferent between a 50/50 lottery over $x$ and $z$ and a 50/50 lottery over $y$ and $z$. Thus, expected utility leaves no room for an individual's tolerance for risk to depend on the type of consequence involved, holding fixed how those consequences are valued under certainty. This interchangeability property is also shared by many popular alternatives to expected utility.

A large empirical and experimental literature suggests that risk attitudes are not fully portable across contexts (see Section 4). For example, Barseghyan, Prince, and Teitelbaum (2011) and Einav, Finkelstein, Pascu, and Cullen (2012) document that risk attitudes inferred from different insurance markets or contractual settings are only imperfectly correlated. These findings cast doubt on the existence of a single measure of risk aversion that can be applied globally. These studies do not, however, separate attitudes toward risk from the utility assigned to outcomes; they use a standard utility-based notion of risk aversion, with expected utility either imposed or used as a benchmark. The examples below show a different form of context dependence, in which lottery preferences change when outcomes are replaced by alternatives that are indifferent under certainty, contrary to the interchangeability property.

**Example 1** (Vacation choice)**.** An individual who contemplates going on a vacation has the following preferences:

$$\$3000 \sim \text{Vacation in Rome}$$
$$\$2950 \sim \text{Vacation in New York}$$
$$\$2800 \sim \text{Vacation in Barcelona}$$

Suppose further that the agent is risk neutral over monetary lotteries. Under expected utility, this implies that

$$\frac{1}{2}(\text{Rome}) + \frac{1}{2}(\text{Barcelona}) \sim \frac{1}{2}(\$3000) + \frac{1}{2}(\$2800) \prec \$2950 \sim \text{New York.}$$

Hence, the agent should prefer going to New York for sure over an equal-chance lottery between Rome and Barcelona.

Suppose, instead, that the agent prefers the lottery between Rome and Barcelona. He explains that he is more tolerant of risk when the objects of uncertainty are urban vacation destinations rather than money. Expected utility cannot accommodate this: because each destination is indifferent to its monetary equivalent, lotteries over destinations must be evaluated exactly as the corresponding lotteries over money.

Evidence from experiments (e.g., Toussaert (2024)) and from real-world companies offering surprise trips suggests that preferences of this kind are both plausible and observable in practice.[1] Moreover, as the following example illustrates, what constitutes a domain may be subjective.

**Example 2** (Vacation choice continued)**.** The agent also considers a vacation in the mountains. His preferences are:

$$\$3000 \sim \text{Vacation in the Rockies}$$
$$\$2800 \sim \text{Vacation in the Alps}$$

He further states that his monetary equivalent of a 50/50 lottery between the Rockies and the Alps is \$2900. His attitude toward risk over mountain vacations coincides with his attitude toward monetary risk, even though his attitude toward risk over urban vacations does not. The relevant distinction is therefore not simply between money and vacations, but between domains revealed by the individual's choices.

[1] For instance, "mystery travel" firms (e.g., Pack Up + Go, Journee, The Vacation Hunt, FlyKube) first elicit travelers' preferences (e.g., trip type, budget) and then handle the entire planning and booking process, revealing the final destination only shortly before departure.

To accommodate these and related examples, a model must permit failures of prize interchangeability. An agent may be indifferent between $x$ and $y$ under certainty, yet prefer a 50/50 lottery over $x$ and $z$ to a 50/50 lottery over $y$ and $z$, depending on the subjective domains to which the prizes belong. More generally, suppose $x \sim y$ and $x' \sim y'$. The agent may prefer the probability-matched lottery over $(x, x')$ to the corresponding lottery over $(y, y')$. Such comparisons form the basis of our notion of greater pure risk aversion (Definition 1); in this case, the agent is more pure risk averse over the y-prizes than over the x-prizes. By matching consequences that are indifferent under certainty and applying the same probabilities, the comparison holds deterministic utility fixed and isolates the individual's attitude toward risk.

We develop a model of domain-specific risk preferences that can accommodate these pure risk attitudes. The model goes beyond expected utility while preserving expected utility within each domain. Risk attitudes may differ across domains, but the departure from the classical model is restricted to comparisons involving prizes from different domains. Section 2.1 introduces our behavioral notion of pure risk aversion, and Section 2.2 presents the pure risk representation.

Our first main result, presented in Section 2.3, is that the relevant domains need not be specified in advance; they can be identified from choice behavior. We first introduce a behavioral relation on prizes, called relatedness. Theorem 1 establishes that two prizes are related if and only if they belong to the same domain in our model. The theorem therefore identifies which prizes are treated as interchangeable and which are treated as categorically distinct. This implies the uniqueness of the relevant domains. Proposition 3 establishes the uniqueness of the remaining parameters.

This identification result also suggests a possible application. Suppose an analyst estimates risk aversion from health-insurance choices and wishes to predict behavior in another context, such as portfolio choice. If the relevant outcomes are related and therefore belong to the same domain, then estimates obtained from health-insurance choices may be used to predict behavior in the new context. If they are not related, then such extrapolation is not justified, and risk aversion for the other context must be estimated separately. The model thus suggests a behavioral criterion for determining when a measure of risk aversion is portable across contexts.

Section 3 contains our second main result, Theorem 2, which provides an axiomatic characterization of the model. Although our representation resembles familiar models in the literature, including a version of second-order expected utility under ambiguity, its characterization requires several axioms specific to our setting. The main task is to identify the domains from choice behavior and impose the appropriate structure on preferences both within and across these subjective domains.

# 2 Model

## 2.1 Pure Risk Aversion

Let $X$ be a finite set of prizes and let $\Delta X$ denote the set of lotteries over $X$.[2] We let $\delta_x$ denote the degenerate lottery that yields prize $x \in X$ for sure. Consider preferences $\succeq$ over $\Delta X$. We impose no structure on $X$, so our comparative notion of pure risk attitude relies only on $\succeq$. Prizes may themselves be bundles of goods, and the domains can partition these bundles without requiring any underlying product structure. Our notion also applies to non-numerical prizes, where expected values cannot be computed in any meaningful way.

We say two sets of prizes $Y \subset X$ and $Z \subset X$ are *comparable* if there exist $y_1, y_2 \in Y$ and $z_1, z_2 \in Z$ such that $y_1 \sim z_1$ and $y_2 \sim z_2$.

**Definition 1.** For comparable $Y$ and $Z$, we say that $\succeq$ is *more pure risk averse over $Y$ than over $Z$* if for all $y_1, y_2 \in Y$, $z_1, z_2 \in Z$, $y_1 \sim z_1$, $y_2 \sim z_2$ and $a \in [0,1]$,

$$a\delta_{y_1} + (1-a)\,\delta_{y_2} \preceq a\delta_{z_1} + (1-a)\,\delta_{z_2}.$$

If $\succeq$ is also more pure risk averse over $Z$ than over $Y$, then we say it is *pure risk neutral* between $Y$ and $Z$.

[2]We restrict to finite $X$ throughout for tractability and to keep the analysis as simple as possible without requiring any topological conditions.

A basic axiom of choice under risk is the weak axiom of degenerate independence (WADI) from Grant, Kajii, and Polak (1992). It is satisfied by many popular models, including expected utility.[3]

**Axiom** (WADI)**.** *For any $x, y \in X$, $p \in \Delta X$, and $a \in [0,1]$, if $x \sim y$ then*

$$a\delta_x + (1-a)\,p \sim a\delta_y + (1-a)\,p.$$

WADI requires that if an individual is indifferent between one certain outcome and another, this indifference is preserved when both outcomes are mixed with any common lottery. WADI immediately rules out domain-specific pure risk attitudes.

**Proposition 1.** *If $\succeq$ satisfies WADI, then $\succeq$ must be pure risk neutral over all comparable domains.*

*Proof.* Suppose $Y, Z \subset X$, $y_1, y_2 \in Y$, $z_1, z_2 \in Z$, $y_1 \sim z_1$, and $y_2 \sim z_2$. By applying WADI twice and using transitivity,

$$a\delta_{y_1} + (1-a)\,\delta_{y_2} \sim a\delta_{z_1} + (1-a)\,\delta_{y_2} \sim a\delta_{z_1} + (1-a)\,\delta_{z_2}.$$

Hence, preferences are pure risk neutral between $Y$ and $Z$. □

Intuitively, if risk attitudes are domain-specific, then the utility values of prizes do not capture the whole picture. An agent may be indifferent between $x$ and $y$, yet risk involving $x$ may be qualitatively different from risk involving $y$. For this reason, prizes with the same utility value need not be interchangeable in risky prospects.

Proposition 1 implies that, to accommodate domain-dependent risk attitudes, one must relax WADI and hence the independence axiom.[4] Our model will depart from expected utility to the extent necessary to capture our richer notion of pure risk aversion.

[3] In particular, it is satisfied by the natural extensions of rank-dependent utility (Quiggin (1982), Yaari (1987)) and disappointment aversion (Gul (1991)) to prizes beyond monetary values.

[4] The independence axiom states that for all $p, q, r \in \Delta X$ and $a \in [0,1]$, $p \succeq q$ implies $ap + (1-a)\,r \succeq aq + (1-a)\,r$.

### 2.2 Pure Risk Representation

We now introduce our functional form. Let $\mathcal{D} = \{D_1, D_2, \ldots, D_n\}$ be a partition of $X$ where each $D_i \subset X$ is a domain. For now, we take $\mathcal{D}$ as given (in Section 2.3, we demonstrate how the domains can be identified from choice behavior). Observe that any lottery $p \in \Delta X$ can be written as

$$p = \sum_i p(D_i)\, q_i$$

where $p(D_i) = \sum_{x \in D_i} p(x)$, and each $q_i \in \Delta D_i$ is the conditional lottery in domain $D_i$ whenever $p(D_i) > 0$. Recall that a function $V : \Delta X \to \mathbb{R}$ represents $\succeq$ when $p \succeq q$ iff $V(p) \geq V(q)$.

**Definition 2.** The relation $\succeq$ has a *pure risk* representation if it is represented by

$$V(p) = \sum_{D_i \in \mathcal{D}} p(D_i)\, \phi_i^{-1}\left( \sum_{x \in D_i} q_i(x)\, \phi_i(u(x)) \right) \tag{1}$$

where $\mathcal{D}$ is a partition of $X$, $u : X \to \mathbb{R}$ is a utility function over prizes, and each $\phi_i : U_i \to \mathbb{R}$ is strictly increasing and continuous, where $U_i \subset \mathbb{R}$ is the smallest interval containing $u(D_i)$.

We identify a pure risk representation with the tuple $(\mathcal{D}, u, \phi)$, where $\phi = (\phi_1, \ldots, \phi_n)$. The inner expression in (1) computes expected utility within domain $D_i$, while $\phi_i^{-1}$ converts it into a certainty equivalent on the common $u$-scale. The overall evaluation is the probability-weighted average of these domain-specific certainty equivalents.

Note that if there is a single domain, i.e., $\mathcal{D} = \{X\}$, then the representation

$$V(p) = \phi^{-1}\left( \sum_x p(x)\phi(u(x)) \right)$$

induces expected utility preferences with Bernoulli index $\phi \circ u$.

**Example 2 continued.** Let $X = \{$Barcelona, Rome, New York, Alps, Rockies, \$2800, \$2900, \$2950, \$3000$\}$ where the two domains are

$$D_1 = \{\text{Alps}, \text{Rockies}, \$2800, \$2900, \$2950, \$3000\}$$

$$D_2 = \{\text{Barcelona}, \text{Rome}, \text{New York}\}$$

Let $p$ be a uniform lottery over $X$, that is, $p = \frac{2}{3}q_1 + \frac{1}{3}q_2$, where $q_1$ and $q_2$ are uniform distributions over $D_1$ and $D_2$, respectively. $V(p)$ then takes the following form:

$$V(p) = \frac{2}{3}\phi_1^{-1}\left(\frac{1}{6}\sum_{x \in D_1} \phi_1 (u(x))\right) + \frac{1}{3}\phi_2^{-1}\left(\frac{1}{3}\sum_{x \in D_2} \phi_2 (u(x))\right).$$

Here the agent treats mountain vacations and monetary payoffs in the same domain, while urban vacations form a distinct domain.

To see how a pure risk representation can generate the preferences in the motivating examples from the introduction, normalize $u(\$2800) = 0$, $u(\$2950) = \frac{3}{4}$, and $u(\$3000) = 1$. Take $\phi_1(t) = t$ and $\phi_2(t) = t^4$, and observe that the monetary and mountain 50/50 lotteries have value $\frac{1}{2}$, corresponding to $u(\$2900)$ (the agent is risk neutral over money), while the urban 50/50 lottery has value $2^{-\frac{1}{4}} > \frac{3}{4}$ and is therefore preferred to New York. In particular, Rome $\sim$ Rockies, Barcelona $\sim$ Alps, but

$$\frac{1}{2}(\text{Rome}) + \frac{1}{2}(\text{Barcelona}) \succ \frac{1}{2}(\text{Rockies}) + \frac{1}{2}(\text{Alps}).$$

The following result confirms that we can interpret $\phi_i$ as capturing pure risk attitudes. Recall that two sets of prizes $Y$ and $Z$ are comparable if there are prizes $y_1, y_2 \in Y$ and prizes $z_1, z_2 \in Z$ such that $y_1 \sim z_1$ and $y_2 \sim z_2$.

**Proposition 2.** *Let $\succeq$ be represented by $(\mathcal{D}, u, \phi)$, and $D_i$ and $D_j$ be comparable. If $\phi_i$ is more concave than $\phi_j$, then $\succeq$ is more pure risk averse over $D_i$ than over $D_j$.*

*Proof.* Let $y_1, y_2 \in D_i$, $z_1, z_2 \in D_j$, $y_1 \sim z_1$, $y_2 \sim z_2$. Let $s = u(y_1) = u(z_1)$ and $t = u(y_2) = u(z_2)$. If $\phi_i$ is more concave than $\phi_j$, then

$$\phi_i^{-1}(a\phi_i(s) + (1-a)\phi_i(t)) \leq \phi_j^{-1}(a\phi_j(s) + (1-a)\phi_j(t))$$

which implies that

$$a\delta_{y_1} + (1-a)\delta_{y_2} \preceq a\delta_{z_1} + (1-a)\delta_{z_2}.$$

Therefore, $\succeq$ is more pure risk averse over $D_i$ than over $D_j$. $\square$

Proposition 2 establishes that the curvature of $\phi_i$ captures pure risk attitudes. This plays the same role as the curvature of Bernoulli utility under expected utility.

Unlike standard Arrow–Pratt comparisons, however, the result does not require prizes to have a numerical or monetary structure. We end this section with several remarks.

**Interpretation of domains.** The domains in our model are defined behaviorally. They capture which prizes the agent treats similarly under risk, without taking a stand on why the partition arises. The partition may reflect some feature of the prizes themselves or of the risky prospect. In this sense, we view the domains as a reduced-form description of how the agent groups prizes under risk.

**Inter-domain risk.** The curvature of $\phi_i$ is also informative about the comparison between intra-domain risk (i.e., risk in that domain) and inter-domain risk (i.e., risk across domains). For example, suppose $x$ and $y$ are in domain $D_i$ while $z$ is in domain $D_j$, where $z \sim x$. In this case, $a\delta_x + (1-a)\,\delta_y$ corresponds to intra-domain risk, i.e., risk in $D_i$, while $a\delta_z + (1-a)\,\delta_y$ corresponds to inter-domain risk, i.e., risk between $D_i$ and $D_j$. If $\phi_i$ is concave, the agent prefers inter-domain risk; if it is convex, the agent prefers intra-domain risk.

**Non-comparable domains.** Our definition of pure risk aversion is based on indifferent prizes in comparable domains. This provides a simple test and is easy to relate to axioms such as WADI. However, it could be that there are no such indifferences, in which case the domains are not comparable. One could extend Definition 1 to lotteries as follows: if for all lotteries $p_1, p_2 \in \Delta Y$, $q_1, q_2 \in \Delta Z$, $p_1 \sim q_1$, $p_2 \sim q_2$, and $a \in [0,1]$ we have $ap_1 + (1-a)\,p_2 \preceq aq_1 + (1-a)\,q_2$, then $\succeq$ is more pure risk averse over $Y$ than over $Z$. This would allow for richer comparisons of pure risk attitudes. It would also strengthen our definition of more pure risk averse so that in Proposition 2, it is both necessary and sufficient for $\phi_i$ to be more concave than $\phi_j$.

**More general representations.** In our model, we relax WADI to accommodate richer pure risk attitudes across domains. At the same time, when faced with risk within a domain or risk between domains, i.e., mixtures where the risk within each domain is fixed, choices satisfy the independence axiom.[5] We view the resulting representation as a tractable benchmark: it preserves expected utility within domains and takes a weighted average of the corresponding domain-specific evaluations. The inde-pendence axiom may also fail within a domain for other reasons, such as Allais-type behavior. A natural extension would allow for such non-expected utility phenomena within domains as well.

[5] In particular, preferences satisfy Intra- and Inter-Domain Independence (Axioms 3 and 4).

## 2.3 Identification

Since the set of domains, i.e., the partition $\mathcal{D} = \{D_1, D_2, \ldots, D_n\}$, is subjective, a natural question is how it can be identified from choice data. We show that certain failures of independence involving only a few prizes can be used to infer how the agent subjectively categorizes prizes into domains. To formalize this idea, we define a behavioral relation $\simeq$ on $X$.

**Definition 3.** For any $x, y \in X$, we say that $x$ and $y$ are *related*, written $x \simeq y$, if for any $z \in X$, all $a \in [0,1]$, and all $p, q \in \Delta\{x, y, z\}$,

$$p \sim q \Rightarrow ap + (1-a)\,\delta_z \sim aq + (1-a)\,\delta_z.$$

The relation $\simeq$ is behaviorally observable from comparisons involving lotteries with at most three prizes. If preferences satisfy expected utility, full independence implies that every pair of prizes is related.

Identification can fail in certain knife-edge cases. We call such representations irregular.

**Definition 4.** $(\mathcal{D}, u, \phi)$ is *irregular* if there exist $i \neq j$, $x \in D_i$ and $y \in D_j$, such that $\phi_i$ and $\phi_j$ are piecewise linear and their kinks, if any, occur only at $u(x)$ or $u(y)$.

A representation is *regular* if it is not irregular. In an irregular representation, the piecewise-linear transformations may allow domains to be merged or separated without changing preferences, so the same preferences can admit pure risk representations with different sets of domains (see Example 3 below). A simple sufficient condition for regularity is that every $\phi_i$ is differentiable and at most one is linear. We restrict attention to regular representations in the results that follow and discuss irregularity immediately afterward.

The following is our main identification result.

**Theorem 1.** *Let $\succeq$ be represented by a regular pure risk representation. Then $x \simeq y$ iff $x$ and $y$ are in the same domain.*

*Proof.* See Appendix A. $\square$

To see why the relation $\simeq$ is necessary, suppose $x$ and $y$ are in the same domain. There are two cases to consider, depending on whether $z$ in Definition 3 is in the same domain as $x$ and $y$. If $z$ is in the same domain, then since independence is satisfied within the domain, the result holds. If $z$ is in a different domain, then mixtures with $z$ do not alter the conditional lottery between $x$ and $y$ in either $p$ or $q$, so again independence holds.[6]

Conversely, if two prizes in distinct domains satisfy $x \simeq y$, Lemma 2 in Appendix A implies that the transformations associated with both domains must be piecewise linear, with possible kinks only at the relevant prize utilities. Regularity excludes precisely these cases.

An immediate implication of Theorem 1 is that the set of domains in a regular pure risk representation is unique, independently of the other parameters.

**Corollary 1.** *Suppose $\succeq$ is represented by regular pure risk representations $(\mathcal{D}, u, \phi)$ and $(\mathcal{C}, v, \varphi)$. Then $\mathcal{D} = \mathcal{C}$.*

Theorem 1 also yields a natural way to compare domains across different agents. We say a partition $\mathcal{D}_1$ is *finer* than another partition $\mathcal{D}_2$ if for every $D \in \mathcal{D}_1$ there exists some $D' \in \mathcal{D}_2$ such that $D \subset D'$.

**Corollary 2.** *Suppose $\succeq_1$ and $\succeq_2$ are represented by regular $(\mathcal{D}_1, u, \phi)$ and $(\mathcal{D}_2, v, \varphi)$, respectively. Then the following are equivalent:*

(1) *$x \simeq_1 y$ implies $x \simeq_2 y$ for all $x, y \in X$*

(2) *$\mathcal{D}_1$ is finer than $\mathcal{D}_2$*

*Proof.* First, suppose (2) is true and consider $x \simeq_1 y$. That means there exists some $D \in \mathcal{D}_1$ such that $x, y \in D$. Since $\mathcal{D}_1$ is finer than $\mathcal{D}_2$, we can find some $D' \in \mathcal{D}_2$ such that $D \subset D'$. This means $x$ and $y$ are in the same domain under $\mathcal{D}_2$ so $x \simeq_2 y$.

[6] See Lemma 1 in Appendix A.

Now suppose (1) is true and consider any $D \in \mathcal{D}_1$. We have that $x \simeq_1 y$ for any $x, y \in D$. Since $x \simeq_2 y$ by (1), $x$ and $y$ must be in the same domain under $\mathcal{D}_2$. Since this is true for all $x, y \in D$, it means $D \subset D'$ for some $D' \in \mathcal{D}_2$ as desired. $\square$

We next discuss what happens when regularity fails. First, note that regularity rules out multiple linear $\phi_i$. This is without loss of generality. Indeed, if two domains are associated with linear functions $\phi_i$ and $\phi_j$, these domains can be merged into a single domain endowed with a linear $\phi$. This observation explains why expected utility admits multiple pure risk representations, one for every partition of $X$, while among regular pure risk representations the only possible partition is $\{X\}$.

Are there other cases beyond expected utility where we have multiple pure risk representations? This can occur only in knife-edge cases in which the relevant transformations are piecewise linear with at most two kinks which occur precisely at the utility values of two specific prizes. The following example provides an illustration.

**Example 3.** Let $X = \left\{x, y, \underline{y}, \bar{y}\right\}$. Suppose $u\left(\underline{y}\right) = 0$, $u(y) = 1$ and $u(x) = u(\bar{y}) = 2$. Let $\mathcal{D} = \{D_1, D_2\}$ where $D_1 = \{x\}$ and $D_2 = \left\{y, \underline{y}, \bar{y}\right\}$. Suppose $\phi_1(t) = t$ and

$$\phi_2(t) = \begin{cases} t & 0 \leq t \leq 1 \\ 2t - 1 & 1 \leq t \leq 2 \end{cases}.$$

Thus, $\phi_2^{-1}$ is piecewise linear with a cutoff at 1. For $p = a\delta_y + (1-a)\left(\lambda\delta_{\bar{y}} + (1-\lambda)\delta_{\underline{y}}\right)$,

$$\begin{aligned} V(p) &= \phi_2^{-1}\left(a\phi_2(u(y)) + (1-a)\left(\lambda\phi_2(u(\bar{y})) + (1-\lambda)\phi_2\left(u\left(\underline{y}\right)\right)\right)\right) \\ &= \phi_2^{-1}(a + (1-a)3\lambda). \end{aligned}$$

If $3\lambda \geq 1$, then $a + (1-a)3\lambda \geq 1$ so

$$\phi_2^{-1}(a + (1-a)3\lambda) = a\phi_2^{-1}(1) + (1-a)\phi_2^{-1}(3\lambda).$$

Symmetric reasoning holds for $3\lambda \leq 1$, so we have

$$\begin{aligned} V(p) &= a\phi_2^{-1}(1) + (1-a)\phi_2^{-1}(3\lambda) \\ &= au(y) + (1-a)\phi_2^{-1}\left(\lambda\phi_2(u(\bar{y})) + (1-\lambda)\phi_2\left(u\left(\underline{y}\right)\right)\right). \end{aligned}$$

This means that the same preferences admit a pure risk representation with partition

$\mathcal{C} = \left\{ \{x, y\}, \left\{ \underline{y}, \bar{y} \right\} \right\}$.

The preference in Example 3 admits multiple pure risk representations. Note that $x$ and $y$ are related (i.e., $x \simeq y$) since they belong to the same domain under $\mathcal{C}$ yet they belong to different domains under $\mathcal{D}$. This illustrates that while any two prizes within the same domain must be related, the converse need not hold: related prizes need not belong to the same domain in every pure risk representation. The non-uniqueness in Example 3 arises precisely because $\phi_2$ has a kink at $u(y)$. We conjecture that a full identification methodology that covers such knife-edge cases would entail procedures that are more involved than our simple $\simeq$ relation.[7]

Having established the uniqueness of domains (Corollary 1), we conclude with a result establishing the uniqueness of the model's remaining parameters.

**Proposition 3.** *Suppose* $\succeq$ *is represented by regular* $(\mathcal{D}, u, \phi)$ *and* $(\mathcal{D}, v, \varphi)$ *where* $|\mathcal{D}| \geq 2$. *Assume* $\phi_i$ *and* $\varphi_i$ *are differentiable with* $\phi_i', \varphi_i' > 0$. *Then*

(1) $u = \alpha v + \beta$ *for some* $\alpha > 0$ *and* $\beta$

(2) $\phi_i(t) = \gamma_i \varphi_i \left( \frac{t-\beta}{\alpha} \right) + \zeta_i$ *for some* $\gamma_i > 0$ *and* $\zeta_i$

*Proof.* See Appendix B. □

To understand Proposition 3, first note that, because preferences satisfy expected utility within each domain $D_i$, the composite index $\phi_i \circ u$ is unique up to a positive affine transformation by standard vNM uniqueness. Suppose first that the best and worst prizes belong to different domains. Mixtures between these prizes are then evaluated linearly on the common $u$-scale. By matching every prize with such a mixture, we can identify $u$ up to a common positive affine transformation. Given this normalization of $u$, vNM uniqueness within each domain identifies each $\phi_i$ up to a positive affine transformation. If instead the best and worst prizes belong to the same domain, this argument no longer applies, because mixtures between them are evaluated through that domain's transformation. We then use a prize from another domain as a bridge, obtaining potentially different affine relationships above and below its utility level. The differentiability and positive-derivative assumptions ensure that these relationships have the same slope, thereby ruling out the knife-edge case illustrated in Example 5 in Appendix B. Finally, the assumption that there are at least two domains excludes the special case of standard expected utility, in which only the composite $\phi \circ u$ is identified.

[7] There is an analogy between identifying domains in our setup and identifying ideal events under ambiguity (Gul and Pesendorfer (2014), Denti and Pomatto (2022)). See the discussion at the end of Section 4.

# 3 Characterization

We next provide an axiomatic characterization of the model. The first two axioms are standard. The first states that $\succeq$ is a preference relation. The second (continuity) is sometimes called solvability in the literature (e.g., Dekel (1986)).

**Axiom 1.** *The relation $\succeq$ is complete and transitive.*

**Axiom 2** (Continuity)**.** *For all $p, q, r \in \Delta X$, if $p \succ r \succ q$, then there is $a \in (0, 1)$ such that $r \sim ap + (1 - a)\, q$.*

We next introduce the axioms governing choices within and across the behaviorally induced domains. The identification result in Section 2.3 shows how domains can be recovered when a pure risk representation exists. Ex ante, however, the existence of such a representation is not guaranteed; for example, the relation $\simeq$ need not be transitive. We therefore construct an induced partition of $X$ as follows. Let $\simeq^*$ denote the transitive closure of $\simeq$, i.e., $x \simeq^* z$ whenever there exists a finite chain $x \simeq y_1 \simeq \cdots \simeq y_n \simeq z$, even if $x \not\simeq z$. We define $\mathcal{D}_{\succeq}$ as the partition of $X$ into equivalence classes under $\simeq^*$. Going forward, we refer to elements of $\mathcal{D}_{\succeq}$ as domains.[8]

The first axiom is the standard independence axiom, restricted to lotteries whose support lies within a single domain.

**Axiom 3** (Intra-Domain Independence)**.** *For every $D_i \in \mathcal{D}_{\succeq}$, all $p, q, r \in \Delta D_i$, and every $a \in [0, 1]$, if $p \succeq q$, then $ap + (1 - a)r \succeq aq + (1 - a)r$.*

[8]If the set of domains is objective and known, then our axiomatic characterization applies directly.

Together with the preceding axioms, Axiom 3 implies that, within each subjective domain, the agent behaves as an expected utility maximizer. As a result, variation in risk attitudes across domains is captured entirely at the inter-domain level.

We now turn to mixtures across domains. Recall that every $p \in \Delta X$ can be written as

$$p = \sum_i p(D_i) \, q_i$$

where $p(D_i) = \sum_{x \in D_i} p(x)$ is the probability assigned to domain $D_i \in \mathcal{D}_{\succeq}$, and $q_i \in \Delta D_i$ is the corresponding conditional lottery whenever $p(D_i) > 0$. Define the induced lottery $\mu_p \in \Delta\left(\bigcup_i \Delta D_i\right)$ by

$$\mu_p := \sum_i p(D_i) \, \delta_{q_i}.$$

That is, $\mu_p$ is the compound lottery in which risk across domains is resolved first, followed by risk within the realized domain.

**Definition 5.** For $p, r \in \Delta X$ and $a \in [0,1]$, the tuple $(p, a, r)$ is a *pure inter-domain mixture* if

$$\mu_{ap+(1-a)r} = a\mu_p + (1-a)\, \mu_r.$$

For example, if $p = bq_i + (1-b)q_j$ and $r = cq_i + (1-c)q_j$ for fixed conditional lotteries $q_i \in \Delta D_i$ and $q_j \in \Delta D_j$, then every mixture of $p$ and $r$ is a pure inter-domain mixture: the conditional lotteries, and hence the risk within each domain, remain fixed, while only the probabilities assigned across domains vary.

**Axiom 4** (Inter-Domain Independence)**.** *For pure inter-domain mixtures* $(p, a, r)$ *and* $(q, a, r)$*, if* $p \succeq q$*, then*

$$ap + (1-a)\, r \succeq aq + (1-a)\, r.$$

Axiom 4 imposes independence only on mixtures that preserve the internal composition of risk within each domain and vary only the probability weights assigned across domains.

Note that each lottery $p \in \Delta X$ can be decomposed into a lottery over domains, $p(D_i)$, followed by a conditional lottery $q_i \in \Delta D_i$ within each domain $D_i$. This decomposition resembles the recursive two-stage framework of Kreps and Porteus (1978). Axiom 3 is analogous to second-stage independence, whereas Axiom 4 is analogous to first-stage independence. The latter analogy is only partial, however, because Axiom 4 applies not to all first-stage mixtures, but only to the more restricted class of pure inter-domain mixtures. An additional axiom is therefore required; no such axiom is needed in the Kreps-Porteus model.[9]

**Axiom 5** (Symmetry)**.** *If $p \sim p'$ and $q \sim q'$ where $p, q \in \Delta D_i$, $p', q' \in \Delta D_j$ and $D_i, D_j \in \mathcal{D}_{\succeq}$, then, for every $a \in [0,1]$,*

$$ap + (1-a)\, q' \sim ap' + (1-a)\, q.$$

Axiom 5 does not require risk attitudes to be identical across domains. It requires only that the aggregation of domain-specific lotteries be invariant to replacing each component by an indifferent lottery from another domain. For example, as illustrated in Example 4 below, it rules out risk attitudes that depend on which domain contains the better of two outcomes.

**Theorem 2.** *$\succeq$ satisfies Axioms 1–5 iff it has a regular pure risk representation.*

*Proof.* See Appendix C. □

We next sketch the proof. Let $\bar{x}, \underline{x} \in X$ denote the best and worst prizes and suppose for simplicity they are in different domains. For any lottery $p$, define $V(p) := \lambda_p$, where

$$p \sim \lambda_p \delta_{\bar{x}} + (1-\lambda_p)\, \delta_{\underline{x}}$$

so $V(\cdot)$ represents $\succeq$. Now, for each domain $D_i$, the axioms guarantee that $\succeq$ on $\Delta D_i$ has some expected utility representation with vNM utility $v_i$. Since $V$ also represents

[9] Despite this analogy and the similarity of the functional form (see the discussion in Section 4), the two models are behaviorally orthogonal: Kreps-Porteus preferences are defined over temporal compound lotteries and capture the agent's attitude toward the timing of uncertainty resolution, whereas the present model is defined over simple atemporal lotteries $\Delta X$ and captures the agent's attitude toward risk that varies with the subjective category of the prizes. An agent can satisfy the axioms of the present paper while being indifferent to the timing of uncertainty resolution, and can satisfy the Kreps-Porteus axioms while exhibiting no domain-specific pure risk aversion.

$\succeq$ on $\Delta D_i$, the uniqueness of expected utility representations ensures that

$$V(q) = \psi_i \left( \sum_{x \in D_i} q(x) v_i(x) \right)$$

for some strictly increasing $\psi_i$. Write $p = \sum_i a_i q_i$ where $q_i \in \Delta D_i$ and define $\tilde{q}_i := \lambda_{q_i} \delta_{\bar{x}} + (1 - \lambda_{q_i}) \delta_{\underline{x}}$. Since $q_i \sim \tilde{q}_i$, Axiom 5 allows us to prove that

$$p \sim \sum_i a_i \tilde{q}_i = \left( \sum_i a_i \lambda_{q_i} \right) \delta_{\bar{x}} + \left( 1 - \left( \sum_i a_i \lambda_{q_i} \right) \right) \delta_{\underline{x}}$$

so

$$\begin{aligned} V(p) &= \sum_i a_i \lambda_{q_i} = \sum_i a_i V(q_i) \\ &= \sum_i a_i \psi_i \left( \sum_{x \in D_i} q_i(x) v_i(x) \right). \end{aligned}$$

Finally, define $u(x) := V(\delta_x) = \psi_i(v_i(x))$ for every $x \in D_i$, and let $\phi_i := \psi_i^{-1}$. Then $v_i(x) = \phi_i(u(x))$, and hence

$$V(p) = \sum_i a_i \phi_i^{-1} \left( \sum_{x \in D_i} q_i(x) \phi_i(u(x)) \right),$$

as desired.

We conclude with an example illustrating the necessity of the symmetry axiom.

**Example 4.** There are four prizes and two domains, $D_1 = \{x_1, y_1\}$ and $D_2 = \{x_2, y_2\}$. For any lottery $p = aq_1 + (1-a) q_2$ where $q_i \in \Delta D_i$, define $t_i = q_i(x_i)$ and

$$V(p) = \begin{cases} at_1^2 + (1-a) t_2^2 & \text{if } t_1 \geq t_2 \\ \sqrt{at_1^4 + (1-a) t_2^4} & \text{if } t_1 < t_2. \end{cases}$$

In Appendix D, we verify that the preferences represented by $V$ satisfy Axioms 1–4 but violate Axiom 5. It follows that they do not have a regular pure risk representation.

# 4 Related literature

A large psychology literature questions whether risk attitudes are stable across domains. Findings from Weber, Blais, and Betz (2002), Hanoch, Johnson, and Wilke (2006), and Frey, Pedroni, Mata, Rieskamp, and Hertwig (2017) challenge the conventional view of risk attitude as a stable characteristic. Our definition of pure risk aversion provides a formal test of what it means for risk aversion to be heterogeneous across domains, thereby offering a theoretical framework within which the empirical findings of the psychology literature can be interpreted and tested. In addition, our model accommodates domain-specific risk attitudes with only a minimal departure from expected utility.

A large empirical literature in economics documents variation in risk attitudes across domains. Barseghyan, Prince, and Teitelbaum (2011) find that households exhibit greater risk aversion in their home deductible choices than in their auto deductible choices, while Einav, Finkelstein, Pascu, and Cullen (2012) document heterogeneity across domains such as health insurance and 401(k) investment decisions. In experiments, DeJarnette (2024) finds that subjects exhibit considerably more risk aversion when allocating Amazon.com credit than when allocating monetarily equivalent goods. Toussaert (2024) conducts a choice experiment involving holiday travel similar to Example 1 and finds that monotonicity violations[10] largely disappear when the destinations in the surprise lotteries are replaced by their monetary valuations. Together, these studies indicate that attitudes toward risk can vary across domains. Our setup formalizes this possibility and provides the flexibility beyond expected utility needed to accommodate it. Moreover, while the empirical literature typically does not separately identify utility and risk attitude, our behavioral test isolates the latter by holding deterministic utility fixed.

Several papers study the relation between the evaluation of certain outcomes and preferences under risk. Grant, Kajii, and Polak (1992) study when preferences over many-good lotteries can be reduced to preferences over money lotteries and sure multivariate outcomes. They show that the Axiom of Degenerate Independence (ADI) is required for this reduction, and argue that accommodating intrinsic risk aversion

[10]That is, some subjects strictly prefer destination A to destination B, but strictly prefer an equal-chance lottery between A and B to receiving A for sure. Our benchmark pure risk model assumes expected utility within a domain, yet, as we remark in Section 2.2, it could be generalized to allow for such violations.

(what we call pure risk aversion) requires relaxing ADI.[11] Our paper picks up where they leave off by developing a model of pure risk aversion, showing how the relevant subjective domains can be identified from choice, and providing an axiomatic characterization. Dyer and Sarin (1982) separate risk attitude from the strength of preference over consequences by introducing a separate measure of the latter. They define risk attitude by comparing von Neumann–Morgenstern utility with this strength-of-preference measure. Their approach relies on an auxiliary primitive, rather than standard revealed preference, and maintains expected utility over lotteries.

A related expected-utility literature studies risk when outcomes have several commodities or attributes. Kihlstrom and Mirman (1974) extend the Arrow–Pratt analysis of risk aversion to multiple commodities. In the context of multi-attribute utility, Maas and Wakker (1994) show that, when riskless preferences admit an additive representation, utility independence of one attribute requires the utility index under risk to be a linear or exponential transformation of the additive value function. Also within the expected utility framework, Alaoui and Penta (2025) propose a decomposition of the Bernoulli utility function into two components: a CARA function, which captures a notion of pure risk attitude, and a function capturing marginal utility of wealth under certainty. When there is more than one domain, their notion of pure risk attitude carries across the domains. These last two papers study how the utility index under risk relates to a representation of preferences under certainty. Our question is different: we study whether prizes assigned the same utility can nevertheless be treated differently under risk, a possibility that requires departing from expected utility.

Ke and Zhang (2025) study multidimensional risk when prizes have an exogenous product structure. Their model permits richer forms of aggregation across dimensions, including recursive hierarchies, and identifies the relevant hierarchy from choice. By contrast, we allow arbitrary subjective partitions of the prize space, including partitions that are not compatible with their product structure, and focus on domain-specific pure risk attitudes. Their unidimensional independence axiom paral-lels our intra-domain independence: expected utility is retained within each primitive dimension, while independence may fail globally. Honda and Sun (2024) allow risk attitudes to depend on lottery-specific context: the utility used to evaluate a lottery varies with the probability assigned to sufficiently bad outcomes. Their model aims to address violations of independence in monetary lotteries. By contrast, in our model the evaluation of risk depends on the domains to which the prizes belong. Our framework also applies to non-numerical prizes and focuses on identifying the relevant domains from choice behavior.

[11] ADI is stronger than WADI. They write that "A fully fledged analysis of preferences that do not satisfy ADI is beyond the scope of this paper..."

Our model is a risk-based counterpart to source-dependent models under ambiguity. The composition of functions in our representation resembles the recursive framework of Kreps and Porteus (1978) and the two-stage structure of models such as Klibanoff, Marinacci, and Mukerji (2005) and Grant, Polak, and Strzalecki (2009). Despite these formal similarities and the parallels between some of the axioms, our model has different behavioral content; in particular, as explained in Footnote 9, it is behaviorally orthogonal to the Kreps-Porteus model. Chew and Sagi (2008) introduce conditional small-world event domains as self-contained collections of comparable events and establish probabilistic sophistication within each such domain, providing a foundation for source-dependent risk attitudes. Cappelli, Cerreia-Vioglio, Maccheroni, Marinacci, and Minardi (2021) allow attitudes toward uncertainty to vary across sources and introduce subjective prices for comparing and aggregating source-specific certainty equivalents. Chew, Kader, and Wang (2023) use multiple mixture operators, each representing a source of risk, to obtain a source-recursive expected utility representation for compound risks. The source structure is part of the setup in the latter two papers, whereas Chew and Sagi (2008) define their event domains behaviorally. Our domains are instead subjective partitions of the prize space, identified from choices among lotteries with objective probabilities.

Related identification questions arise in models where the decision maker groups states or events. Burkovskaya (2022) axiomatizes a subjective aggregation of states and recovers the resulting small worlds, beliefs, and utility from asset demand. As in our model, the partition is identified from behavior rather than specified in advance, but hers is over states and ours is over prizes. Gul and Pesendorfer (2014) behav-iorally define ideal events, for which both the event and its complement satisfy a version of the sure-thing principle. Denti and Pomatto (2022) recover from preferences a subjective statistical model and the remaining components of a smooth-ambiguity representation. Our identification result is analogous in spirit: the relation $\simeq$ recovers from choice the subjective domains that determine which prizes are treated as interchangeable. This connection may be useful in both directions. Methods used to identify ideal events may help identify domains in irregular pure risk representations, while our approach may, in turn, help identify ideal events under ambiguity.

# Appendix

## A Proof of Theorem 1

We first prove two lemmas. To simplify notation, we write

$$paq := ap + (1-a)\,q.$$

In the appendix, we identify a prize with the corresponding degenerate lottery when no confusion can arise. Accordingly, we use $xay$ and $a\delta_x + (1-a)\delta_y$ interchangeably.

**Lemma 1.** Suppose $\succeq$ is represented by $(\mathcal{D}, u, \phi)$. If $x$ and $y$ are in the same domain, then $x \simeq y$.

*Proof.* Let $x, y \in D_i \in \mathcal{D}$. If $z \in D_i$, then the result follows because preferences are expected utility within each domain. Thus, assume $z \notin D_i$. Let $p \sim q$ where $p, q \in \Delta\{x, y, z\}$. Note that we can write $p = zap'$ and $q = zbq'$, where $p' = x\lambda y$ and $q' = x\gamma y$. Since $p \sim q$,

$$au\,(z) + (1-a)\,V_i\,(p') = bu\,(z) + (1-b)\,V_i\,(q')\,,$$

where

$$V_i\,(r) = \phi_i^{-1}\left(\sum_s r(s)\phi_i\,(u\,(s))\right).$$

This implies that

$$(1-\kappa)\,u\,(z) + \kappa\,(au\,(z) + (1-a)\,V_i\,(p')) = (1-\kappa)\,u\,(z) + \kappa\,(bu\,(z) + (1-b)\,V_i\,(q'))$$

for any $\kappa \in [0,1]$, and thus $p\kappa z \sim q\kappa z$ as desired. $\square$

**Lemma 2.** Let $\succeq$ be represented by $(\mathcal{D}, u, \phi)$. Suppose $x \simeq y$ where $x \in D_i$ and $y \in D_j$, for $i \neq j$. Let $y' \in D_j$.

(1) If $y' \succeq y \succeq x$ or $x \succeq y' \succeq y$, then $\phi_j$ is linear on $[u\,(y)\,, u\,(y')]$.

(2) If $y \succeq y' \succeq x$ or $x \succeq y \succeq y'$, then $\phi_j$ is linear on $[u\,(y')\,, u\,(y)]$.

(3) If $y' \succ x \succ y$, then $\phi_j$ is linear on $[u\,(y)\,, u\,(x)]$ and $[u\,(x)\,, u\,(y')]$.

(4) If $y \succ x \succ y'$, then $\phi_j$ is linear on $[u\,(y')\,, u\,(x)]$ and $[u\,(x)\,, u\,(y)]$.

*Proof.* We first prove (1). First, suppose $y' \succeq y \succeq x$ so, by continuity, $y'\lambda x \sim y$ for some $\lambda \in [0,1]$. Therefore,

$$\lambda u(y') + (1-\lambda) u(x) = u(y).$$

Since $x \simeq y$, this means that $(y'\lambda x) ay' \sim yay'$ or

$$a(\lambda u(y') + (1-\lambda) u(x)) + (1-a) u(y') = \phi_j^{-1}(a\phi_j(u(y)) + (1-a)\phi_j(u(y')))$$
$$\phi_j(au(y) + (1-a) u(y')) = a\phi_j(u(y)) + (1-a)\phi_j(u(y'))$$

for all $a \in [0,1]$. This shows that $\phi_j$ is linear between $u(y)$ and $u(y')$, as desired.

Now suppose $x \succeq y' \succeq y$. Let $p = y'ay$ for some $a \in (0,1)$ and suppose $p \sim q = xby$ for some $b \in [0,1]$. Note that if $b = 1$, then $y'ay \sim x \succeq y'$ so $y' \sim y$ in which case the result is trivial. On the other hand, if $b = 0$, then $y'ay \sim y$ so again $y' \sim y$ and the result is trivial. Assume then $b \in (0,1)$ and define $k := \frac{a}{b} > a$. Consider the Machina-Marschak triangle where $x, y', y$ correspond to the points $(0,k), (1,1)$, and $(0,0)$, respectively (see Figure A.1). Thus, $p$ and $q$ correspond to $(a,a)$ and $(0,a)$, respectively. Let $r$ be the lottery corresponding to the intersection of the line from $(0,k)$ to $(a,a)$ and the line from $(0,a)$ to $(1,1)$, that is

$$\left(\frac{a(k-a)}{k-a^2}, \frac{a(k-a+k(1-a))}{k-a^2}\right).$$

Define

$$f(a) := \frac{a(k-a+k(1-a))}{k-a^2}.$$

Let $p_1$ be the lottery corresponding to $(f(a), f(a))$ and $q_1$ be the lottery corresponding to $(0, f(a))$. Recall that $p \sim q$. Since $y$ and $y'$ belong to the same domain, Lemma 1 implies that $y \simeq y'$. Applying $y \simeq y'$ with $x$ as the third prize gives $r \sim q_1$, while applying $x \simeq y$ with $y'$ as the third prize gives $r \sim p_1$. Hence, $p_1 \sim q_1$. If we let $p_n$ and $q_n$ be the lotteries corresponding to $(f^n(a), f^n(a))$ and $(0, f^n(a))$ respectively, then by repeating the above argument, $p_n \sim q_n$.

We now show that $f(a)$ is bounded by $\min\{k, 1\}$. Note that

$$1 - f(a) = \frac{(1-a)^2 k}{k-a^2} > 0$$

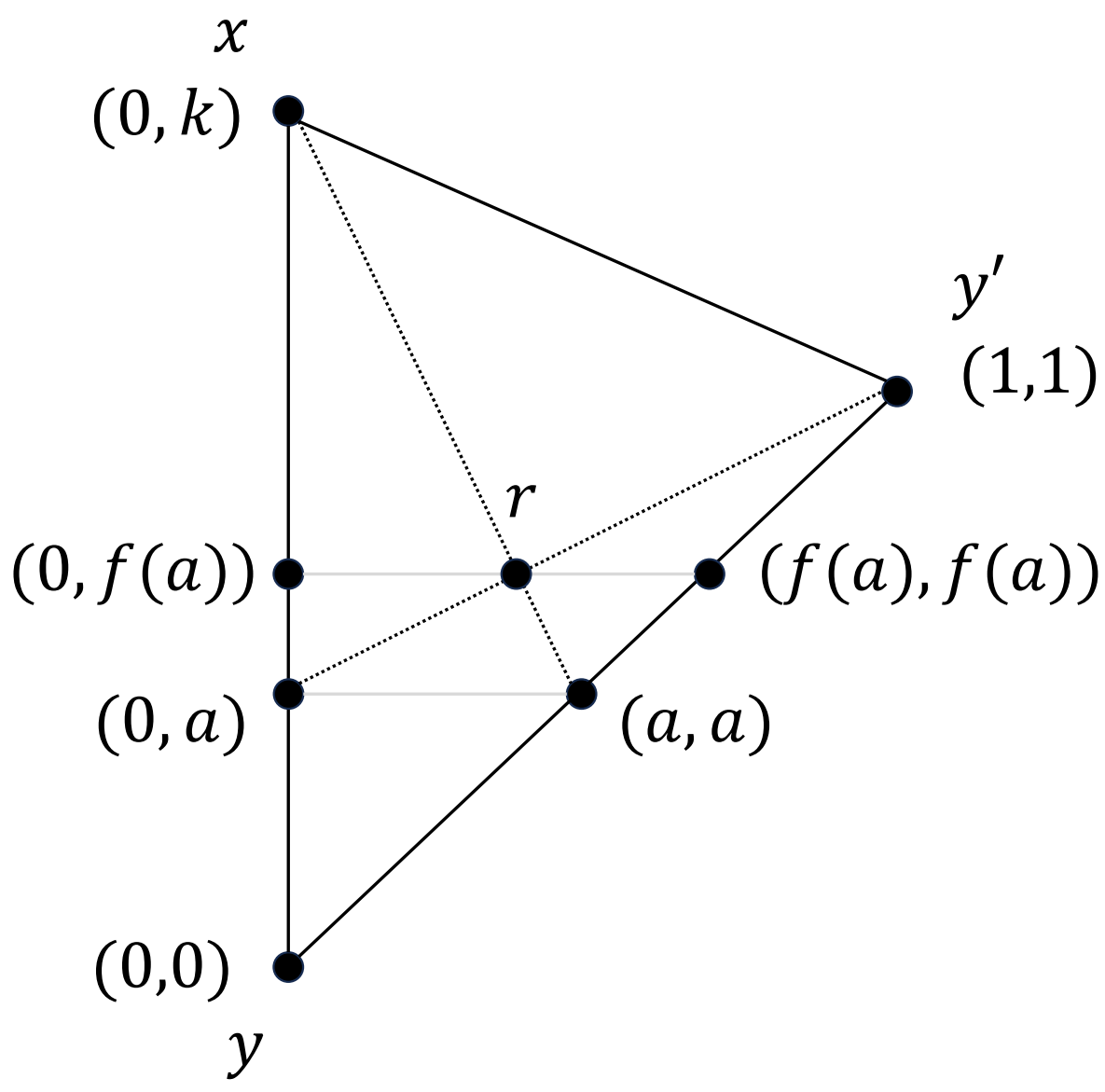

Figure A.1: Proof for the $x \succeq y' \succeq y$ case.

and

$$k - f(a) = \frac{(k-a)^2}{k - a^2} > 0.$$

To see that $f$ always returns a strictly larger number, note that

$$f(a) - a = \frac{a(k-a)(1-a)}{k-a^2} > 0.$$

Therefore, if we let $a_n := f^n(a)$, then $a_n$ is strictly increasing and bounded so it must converge to some $a^* = \lim_n a_n$. Since $f$ is continuous,

$$f(a^*) = f\left(\lim_n a_n\right) = \lim_n f(a_n) = \lim_n a_{n+1} = a^*.$$

It is easy to check that $k$ and 1 are the fixed points of $f$, so $a_n \to \min\{k, 1\}$. Now, suppose $b > a$ so $k < 1$ and let $p^*$ be the lottery corresponding to $(k, k)$. By continuity, we have $x \sim p^* = y'\lambda y$ for some $\lambda \in (0,1)$. This implies $y' \sim y \sim x$ in which case the result is trivial. Assume then $b \leq a$, so $k \geq 1$. Note that $x\frac{b}{a}y$ is the lottery corresponding to $(0,1)$ so by continuity, $x\frac{b}{a}y \sim y'$.

Now, let $x\lambda y \sim y'$ for some $\lambda \in [0,1]$, so that

$$\lambda u(x) + (1-\lambda) u(y) = u(y').$$

Since for any $a \in (0,1)$, $x\frac{b}{a}y \sim y'$, $\lambda = \frac{b}{a}$. Thus, we have

$$y'ay \sim xby = x(\lambda a)y = (x\lambda y)ay$$

so

$$\begin{aligned}\phi_j^{-1}(a\phi_j(u(y')) + (1-a)\phi_j(u(y))) &= a(\lambda u(x) + (1-\lambda)u(y)) + (1-a)u(y) \\ a\phi_j(u(y')) + (1-a)\phi_j(u(y)) &= \phi_j(au(y') + (1-a)u(y)).\end{aligned}$$

Therefore, $\phi_j$ is linear between $u(y)$ and $u(y')$ as desired. This proves (1).

For (2), if $y \succeq y' \succeq x$, then the result holds from the same argument as the second part of (1) above. If $x \succeq y \succeq y'$, then the result holds from the same argument as the first part of (1) above.

For (3), consider $\lambda \in (0,1)$ such that $x \sim y'\lambda y$. That $\phi_j$ is linear on $[u(x), u(y')]$ follows from the first part of (1) above. That $\phi_j$ is linear on $[u(y), u(x)]$ follows from the second part of (1) above. Finally, the case for (4) is symmetric to (3). □

Given Lemma 1, to prove Theorem 1 all we need to show is that if $x \simeq y$, then $x$ and $y$ must be in the same domain. If not, then $x \simeq y$ with $x \in D_i$, $y \in D_j$ for $i \neq j$. By Lemma 2, $\phi_j$ is piecewise linear on $U_j$, with possible kinks only at $u(x)$ and $u(y)$. Applying the same lemma with $x$ and $y$ interchanged shows that the same is true of $\phi_i$ on $U_i$. Hence, $(\mathcal{D}, u, \phi)$ is irregular, yielding a contradiction.

# B Proof of Proposition 3

Note that by vNM uniqueness within each domain $D_i$, we have

$$\phi_i(u(x)) = \gamma_i \varphi_i(v(x)) + \zeta_i$$

for all $i$, where $\gamma_i > 0$. Let $\bar{x}, \underline{x} \in X$ denote the best and worst prizes in $X$.

First, assume that $\bar{x}$ and $\underline{x}$ are in different domains (recall that $|\mathcal{D}| \geq 2$). For any

$x \in D_i$, we can find $\lambda$ such that $\delta_x \sim \bar{x}\lambda\underline{x}$. We have

$$u(x) = \lambda u(\bar{x}) + (1-\lambda) u(\underline{x})$$
$$v(x) = \lambda v(\bar{x}) + (1-\lambda) v(\underline{x}).$$

This implies that

$$\begin{aligned} u(x) &= \frac{v(x) - v(\underline{x})}{v(\bar{x}) - v(\underline{x})} u(\bar{x}) + \frac{v(\bar{x}) - v(x)}{v(\bar{x}) - v(\underline{x})} u(\underline{x}) \\ &= \frac{u(\bar{x}) - u(\underline{x})}{v(\bar{x}) - v(\underline{x})} v(x) + \frac{v(\bar{x}) u(\underline{x}) - v(\underline{x}) u(\bar{x})}{v(\bar{x}) - v(\underline{x})} \\ &= \alpha v(x) + \beta \end{aligned}$$

for some $\alpha > 0$ and $\beta$, proving (1). This further implies that

$$\phi_i(u(x)) = \gamma_i \varphi_i\left(\frac{u(x) - \beta}{\alpha}\right) + \zeta_i.$$

Now, let $x^*, x_*$ denote the best and worst prizes in $D_i$. For any $a \in [0,1]$, we can find some $\lambda$ such that $x^* a x_* \sim \bar{x}\lambda\underline{x}$. This implies that

$$a\phi_i(u(x^*)) + (1-a)\phi_i(u(x_*)) = \phi_i(\lambda u(\bar{x}) + (1-\lambda) u(\underline{x}))$$
$$a\varphi_i(v(x^*)) + (1-a)\varphi_i(v(x_*)) = \varphi_i(\lambda v(\bar{x}) + (1-\lambda) v(\underline{x})).$$

Thus,

$$\begin{aligned} \phi_i(\lambda u(\bar{x}) + (1-\lambda) u(\underline{x})) &= a(\gamma_i \varphi_i(v(x^*)) + \zeta_i) + (1-a)(\gamma_i \varphi_i(v(x_*)) + \zeta_i) \\ &= \gamma_i (a\varphi_i(v(x^*)) + (1-a)\varphi_i(v(x_*))) + \zeta_i \\ &= \gamma_i (\varphi_i(\lambda v(\bar{x}) + (1-\lambda) v(\underline{x}))) + \zeta_i \\ &= \gamma_i \varphi_i\left(\frac{(\lambda u(\bar{x}) + (1-\lambda) u(\underline{x})) - \beta}{\alpha}\right) + \zeta_i. \end{aligned}$$

This implies that for any $t = \lambda u(\bar{x}) + (1-\lambda) u(\underline{x})$,

$$\phi_i(t) = \gamma_i \varphi_i\left(\frac{t - \beta}{\alpha}\right) + \zeta_i$$

showing (2) as desired.

Now, suppose $\bar{x}$ and $\underline{x}$ belong to the same domain, denoted $D_1$. Let $y_*$ be the worst prize in $X \setminus D_1$, and suppose $y_* \in D_2$ (again, this is possible as there are at least two domains). We may further assume that $\bar{x} \succ y_* \succ \underline{x}$ without loss of generality (if there is indifference, then just take $y_*$ as the best or worst prize). Now, for any $x \in D_i$ if $x \succeq y_*$, we can find $\lambda$ such that $\delta_x \sim \bar{x}\lambda y_*$ so

$$u(x) = \lambda u(\bar{x}) + (1-\lambda) u(y_*)$$
$$v(x) = \lambda v(\bar{x}) + (1-\lambda) v(y_*).$$

By the same argument as above,

$$\begin{aligned} u(x) &= \frac{u(\bar{x}) - u(y_*)}{v(\bar{x}) - v(y_*)} v(x) + \frac{v(\bar{x}) u(y_*) - v(y_*) u(\bar{x})}{v(\bar{x}) - v(y_*)} \\ &= \alpha_1 v(x) + \beta_1. \end{aligned}$$

On the other hand, if $x \preceq y_*$, then we can find $\lambda$ such that $\delta_x \sim y_* \lambda \underline{x}$ so by the same argument,

$$\begin{aligned} u(x) &= \frac{u(y_*) - u(\underline{x})}{v(y_*) - v(\underline{x})} v(x) + \frac{v(y_*) u(\underline{x}) - v(\underline{x}) u(y_*)}{v(y_*) - v(\underline{x})} \\ &= \alpha_2 v(x) + \beta_2. \end{aligned}$$

It follows that

$$u(x) = \begin{cases} \alpha_1 v(x) + \beta_1 & \text{if } u(x) \geq u(y_*) \\ \alpha_2 v(x) + \beta_2 & \text{if } u(x) < u(y_*). \end{cases}$$

Moreover,

$$\alpha_1 v(y_*) + \beta_1 = u(y_*) = \alpha_2 v(y_*) + \beta_2.$$

Now, let $x^*, x_*$ denote the best and worst prizes in $D_i$. For any $a \in [0,1]$ such that $x^* a x_* \succeq y_*$, we can find some $\lambda$ such that $x^* a x_* \sim \bar{x}\lambda y_*$. Then

$$a\phi_i(u(x^*)) + (1-a)\phi_i(u(x_*)) = \phi_i(\lambda u(\bar{x}) + (1-\lambda) u(y_*))$$
$$a\varphi_i(v(x^*)) + (1-a)\varphi_i(v(x_*)) = \varphi_i(\lambda v(\bar{x}) + (1-\lambda) v(y_*)).$$

Since $\phi_i(u(x)) = \gamma_i \varphi_i(v(x)) + \zeta_i$,

$$\begin{aligned}
\phi_i(\lambda u(\bar{x}) + (1-\lambda) u(y_*)) &= \gamma_i (a\varphi_i(v(x^*)) + (1-a)\varphi_i(v(x_*))) + \zeta_i \\
&= \gamma_i \varphi_i(\lambda v(\bar{x}) + (1-\lambda) v(y_*)) + \zeta_i \\
&= \gamma_i \varphi_i\left(\frac{\lambda u(\bar{x}) + (1-\lambda) u(y_*) - \beta_1}{\alpha_1}\right) + \zeta_i.
\end{aligned}$$

Hence for any such $t = \lambda u(\bar{x}) + (1-\lambda) u(y_*) \geq u(y_*)$, we have

$$\phi_i(t) = \gamma_i \varphi_i\left(\frac{t-\beta_1}{\alpha_1}\right) + \zeta_i.$$

By symmetric reasoning, we have for any such $t = \lambda u(\underline{x}) + (1-\lambda) u(y_*) \leq u(y_*)$,

$$\phi_i(t) = \gamma_i \varphi_i\left(\frac{t-\beta_2}{\alpha_2}\right) + \zeta_i.$$

Summarizing, for all $t \in U_i$,

$$\phi_i(t) = \begin{cases} \gamma_i \varphi_i\left(\frac{t-\beta_1}{\alpha_1}\right) + \zeta_i & \text{if } t \geq u(y_*) \\ \gamma_i \varphi_i\left(\frac{t-\beta_2}{\alpha_2}\right) + \zeta_i & \text{if } t < u(y_*). \end{cases}$$

Since $\phi_i$ and $\varphi_i$ are differentiable with $\varphi_i' > 0$,

$$\phi_i'(u(y_*)) = \gamma_i \varphi_i'\left(\frac{u(y_*) - \beta_1}{\alpha_1}\right)\frac{1}{\alpha_1} = \gamma_i \varphi_i'(v(y_*))\frac{1}{\alpha_1} = \gamma_i \varphi_i'(v(y_*))\frac{1}{\alpha_2}.$$

Thus, $\alpha_1 = \alpha_2$ which then implies $\beta_1 = \beta_2$ as desired. This concludes the proof of Proposition 3.

Finally, the example below demonstrates that the nonzero derivative assumption is necessary. Note that this applies only when the best and worst prizes are in the same domain.

**Example 5.** Let $X = \{\underline{x}, \bar{x}, y\}$ and $\mathcal{D} = \{D_1, D_2\}$ where $D_1 = \{\underline{x}, \bar{x}\}$ and $D_2 = \{y\}$. Consider the representation $(\mathcal{D}, u, \phi)$ where $u(\bar{x}) = 1$, $u(\underline{x}) = -1$, $u(y) = 0$, $\phi_1(t) = t^3$ and $\phi_2(t) = t$. For any lottery $p = \lambda\left(a\delta_{\bar{x}} + (1-a)\delta_{\underline{x}}\right) + (1-\lambda)\delta_y$,

$$\begin{aligned}
V(p) &= \lambda \phi_1^{-1}(a\phi_1(1) + (1-a)\phi_1(-1)) \\
&= \lambda (2a-1)^{\frac{1}{3}}.
\end{aligned}$$

Now, consider the representation $(\mathcal{D}, v, \varphi)$ where $v(\bar{x}) = 2$, $v(\underline{x}) = -1$, $v(y) = 0$,

$\varphi_2(t) = t$, but

$$\varphi_1(t) = \begin{cases} \left(\frac{t}{2}\right)^3 & \text{if } t \geq 0 \\ t^3 & \text{if } t < 0. \end{cases}$$

Note that $\varphi_1$ is differentiable and

$$\varphi_1^{-1}(s) = \begin{cases} 2s^{\frac{1}{3}} & \text{if } s \geq 0 \\ s^{\frac{1}{3}} & \text{if } s < 0. \end{cases}$$

Under this representation

$$\begin{aligned} \hat{V}(p) &= \lambda \varphi_1^{-1}\left(a\varphi_1(2) + (1-a)\varphi_1(-1)\right) \\ &= \lambda \varphi_1^{-1}(2a-1) \\ &= \begin{cases} 2\lambda(2a-1)^{\frac{1}{3}} & \text{if } 2a-1 \geq 0 \\ \lambda(2a-1)^{\frac{1}{3}} & \text{if } 2a-1 < 0. \end{cases} \end{aligned}$$

Since $\hat{V}$ is a monotone transformation of $V$, $(\mathcal{D}, v, \varphi)$ represents the same preferences as $(\mathcal{D}, u, \phi)$.

# C Proof of Theorem 2

Recall that $\mathcal{D}_{\succeq}$ is the partition generated by the transitive closure of $\simeq$. We first show that regularity is characterized by $\mathcal{D} = \mathcal{D}_{\succeq}$.

**Proposition 4.** *Suppose $\succeq$ is represented by $(\mathcal{D}, u, \phi)$. Then $(\mathcal{D}, u, \phi)$ is regular iff $\mathcal{D} = \mathcal{D}_{\succeq}$.*

*Proof.* Note that if $(\mathcal{D}, u, \phi)$ is regular, we have $x, y \in D$ iff $x \simeq y$ from Theorem 1. That means $\simeq$ is transitive so $\mathcal{D} = \mathcal{D}_{\succeq}$.

Now, suppose $\mathcal{D} = \mathcal{D}_{\succeq}$ but $(\mathcal{D}, u, \phi)$ is not regular, so we can find $x \in D_1 \in \mathcal{D}_{\succeq}$, $y \in D_2 \in \mathcal{D}_{\succeq}$ such that $\phi_1$ and $\phi_2$ are piecewise linear on $U_1$ and $U_2$ with at most two kinks at $u(x)$ and $u(y)$. We will show that $x \simeq y$. Let $x \simeq_z y$ for any $z \in X$ to denote that for any $p, q \in \Delta\{x, y, z\}$ and $a \in [0, 1]$,

$$p \sim q \Rightarrow ap + (1-a)\,\delta_z \sim aq + (1-a)\,\delta_z.$$

Clearly, if $z \notin D_1 \cup D_2$, then $x \simeq_z y$ as $x, y$ and $z$ are all in distinct domains. Thus, we only need to show $x \simeq_z y$ for $z \in D_1 \cup D_2$.

First, suppose $z \in D_1$. Now for any $p \in \Delta\{x, y, z\}$ where $p(\{x, z\}) > 0$,

$$V(p) = p(\{x, z\})\,\phi_1^{-1}\left(\frac{p(x)}{p(\{x, z\})}\phi_1(u(x)) + \frac{p(z)}{p(\{x, z\})}\phi_1(u(z))\right) + p(y)\,u(y).$$

If $u(y)$ is not strictly between $u(x)$ and $u(z)$, then $\phi_1$ is linear between $u(x)$ and $u(z)$ so

$$V(p) = p(x)\,u(x) + p(z)\,u(z) + p(y)\,u(y).$$

Now, suppose $\phi_1$ has a kink at $u(y)$ that is strictly between $u(x)$ and $u(z)$. Note that if

$$\frac{p(x)}{p(\{x, z\})}\phi_1(u(x)) + \frac{p(z)}{p(\{x, z\})}\phi_1(u(z)) \geq \phi_1(u(y))$$

then, since $\phi_1$ is linear on $[u(y), \max\{u(x), u(z)\}]$, we have

$$V(p) = \phi_1^{-1}(p(x)\,\phi_1(u(x)) + p(z)\,\phi_1(u(z)) + p(y)\,\phi_1(u(y))).$$

The case in which

$$\frac{p(x)}{p(\{x, z\})}\phi_1(u(x)) + \frac{p(z)}{p(\{x, z\})}\phi_1(u(z)) \leq \phi_1(u(y))$$

is symmetric, so $x \simeq_z y$.

If $z \in D_2$, then the argument is symmetric so $x \simeq_z y$ for all $z \in X$. But this proves that $x \simeq y$ where $x \in D_1 \in \mathcal{D}_{\succeq}$ and $y \in D_2 \in \mathcal{D}_{\succeq}$. This means $D_1 = D_2$ yielding a contradiction. $\square$

We now show sufficiency in Theorem 2. We first prove several lemmas.

**Lemma 3.** For any $x, y \in X$, preferences on $\Delta\{x, y\}$ have an expected utility representation.

*Proof.* If $x$ and $y$ are in the same domain, then Axioms 1, 3, and 2 ensure preferences on $\Delta\{x, y\}$ have an expected utility representation. If not, then the result holds by

Axioms 1, 4, and 2. □

**Lemma 4.** If $x \succ y$ and $x \succeq p \succeq y$, then there exists a unique $a \in [0,1]$ such that $p \sim xay$.

*Proof.* Note that existence is ensured by Axiom 2 so we just need to prove uniqueness. Suppose otherwise, so $xay \sim xby$ for $a > b$. By Lemma 3, preferences on $\Delta\{x, y\}$ have an expected utility representation so $xay \succ xby$ yielding a contradiction. □

**Lemma 5.** Let $p = \sum_i a_i q_i$ where $q_i \in \Delta D_i$ and suppose $q_i \sim x\lambda_i y$ where $x$ and $y$ are not in the same domain. Then

$$p \sim \sum_i a_i (x\lambda_i y).$$

*Proof.* Without loss of generality, assume $x \in D_1$ and $y \in D_2$ and $x \succeq y$. Note that if $x \sim y$, then by Lemma 3, $q_i \sim x$ for all $q_i$. The conclusion then follows from repeated application of Axiom 4. Thus, assume $x \succ y$.

We first consider the case where

$$p = q_1 a q_2.$$

Since $q_1 \sim x\lambda_1 y$, it must be that $x \succeq q_1 \succeq y$ by Lemma 3. First, suppose $q_1 \succeq q_2$ so by Axiom 2, $q_1 \sim xbq_2$ for some $b \in [0,1]$. By Axiom 4, we have

$$p = q_1 a q_2 \sim (xbq_2)\, aq_2 = x\,(ab)\, q_2.$$

Since $q_2 \sim x\lambda_2 y$, by Axiom 4 again,

$$x\,(ab)\, q_2 \sim x\,(ab)\,(x\lambda_2 y) = x\,(ab + (1 - ab)\,\lambda_2)\, y.$$

Moreover, by Axiom 4

$$q_1 \sim xbq_2 \sim xb\,(x\lambda_2 y) = x\,(b + (1-b)\,\lambda_2)\, y.$$

Since $q_1 \sim x\lambda_1 y$, this means that

$$\lambda_1 = b + (1-b)\,\lambda_2.$$

Thus,

$$\begin{aligned} a\lambda_1 + (1-a)\lambda_2 &= a(b + (1-b)\lambda_2) + (1-a)\lambda_2 \\ &= ab + (1-ab)\lambda_2 \end{aligned}$$

so

$$p \sim x(ab + (1-ab)\lambda_2)y = (x\lambda_1 y)a(x\lambda_2 y)$$

as desired. Now, suppose $q_1 \prec q_2$ so $x \succeq q_2 \succ q_1 \succeq y$. By Axiom 2, we can find $q_i' \in \Delta D_i$ such that $q_1' \sim q_2 \sim x\lambda_2 y$ and $q_2' \sim q_1 \sim x\lambda_1 y$. By Axiom 5,

$$p = q_1 a q_2 \sim q_1'(1-a)q_2'.$$

Since $q_1' \sim x\lambda_2 y$, $q_2' \sim x\lambda_1 y$ and $q_1' \succ q_2'$, the same argument as above yields

$$p \sim (x\lambda_2 y)(1-a)(x\lambda_1 y) = (x\lambda_1 y)a(x\lambda_2 y)$$

as desired. This proves the result for the case where $p = q_1 a q_2$.

Next, consider the case where

$$p = \sum_{i>2} a_i q_i.$$

Since $q_i \sim x\lambda_i y$ and $x, y \notin D_i$ for $i > 2$, we can apply Axiom 4 iteratively and obtain

$$p \sim \sum_{i>2} a_i (x\lambda_i y)$$

as desired.

Finally, consider the general case where

$$p = \gamma q + (1-\gamma) r$$

where $q = q_1 b q_2$ and $r = \sum_{i>2} c_i q_i$. From above, we have

$$\begin{aligned} q &\sim (x\lambda_1 y)b(x\lambda_2 y) \\ r &\sim \sum_{i>2} c_i (x\lambda_i y). \end{aligned}$$

Applying Axiom 4, we have

$$p = q\gamma r \sim [(x\lambda_1 y)b(x\lambda_2 y)]\gamma r \sim [(x\lambda_1 y)b(x\lambda_2 y)]\gamma\left(\sum_{i>2} c_i (x\lambda_i y)\right)$$

as desired. $\square$

**Lemma 6.** If $\bar{z}$ and $\underline{z}$ are the best and worst prizes in $Z \subset X$, then $\bar{z} \succeq p \succeq \underline{z}$ for all $p \in \Delta Z$.

*Proof.* Define $Z_i = Z \cap D_i$ so all the non-empty $Z_i$ form a partition of $Z$ and we can write

$$p = \sum_i a_i q_i$$

where $q_i \in \Delta Z_i$. Let $\bar{z}_i$ and $\underline{z}_i$ be the best and worst prizes in $Z_i$. Since $\succeq$ on $\Delta Z_i$ has an expected utility representation, $\bar{z} \succeq \bar{z}_i \succeq q_i \succeq \underline{z}_i \succeq \underline{z}$. By Axiom 2, $q_i \sim \bar{z}\lambda_i\underline{z}$ for some $\lambda_i$. If $\bar{z}$ and $\underline{z}$ are not in the same domain, then by Lemma 5

$$p \sim \sum_i a_i \left(\bar{z}\lambda_i\underline{z}\right) = \bar{z}\left(\sum_i a_i\lambda_i\right)\underline{z}$$

so $\bar{z} \succeq p \succeq \underline{z}$ by Lemma 3. Now, suppose $\bar{z}$ and $\underline{z}$ are in the same domain, and without loss, assume $\bar{z}, \underline{z} \in Z_1$. Let $\bar{y}$ and $\underline{y}$ be the best and worst prizes in $Z \backslash Z_1$ (if there is only one domain, then the result follows from standard expected utility arguments). If $q_1 \succeq \underline{y}$, then $\bar{z} \succeq q_i \succeq \underline{y}$ for all $i$ so the result holds by the same argument above as $\bar{z}$ and $\underline{y}$ are not in the same domain. If $q_1 \prec \underline{y}$, then $\bar{y} \succeq q_i \succeq \underline{z}$ for all $i$ so the result holds again as $\bar{y}$ and $\underline{z}$ are not in the same domain. This concludes the proof. $\square$

**Lemma 7.** Suppose $z \succeq x \succ y \succeq w$ where $z, w \in D_i$ and $x, y \notin D_i$. Let $x \sim p$ and $y \sim q$ where $p \in \Delta\{z, y\}$, $q \in \Delta\{x, w\}$. Then for every $a \in [0,1]$,

$$ax + (1-a)\,q \sim ap + (1-a)\,y.$$

*Proof.* First, suppose $x$ and $y$ are in different domains. By Axiom 4,

$$ax + (1-a)\,q \sim ax + (1-a)\,y \sim ap + (1-a)\,y$$

as desired. Now, suppose $x$ and $y$ are in the same domain. Let $\hat{p} \sim p \sim x$ and

$\hat{q} \sim q \sim y$ for $\hat{p}, \hat{q} \in \Delta \{z, w\}$. We thus have

$$\begin{aligned} ax + (1-a) q &\sim ax + (1-a) \hat{q} \\ &\sim a\hat{p} + (1-a) y \\ &\sim ap + (1-a) y \end{aligned}$$

where the first and third indifferences follow from Axiom 4 while the second indifference follows from Axiom 5. $\square$

We now prove sufficiency. Note that if there is a single domain, then Axioms 1-3 establish an expected utility representation. Thus, assume there are at least two domains. Let $\bar{x}$ be the best prize in $X$ and assume $\bar{x} \in D_1$ without loss. Let $x_*$ be the worst prize in $D_1$. Let $\bar{y}$ and $\underline{y}$ be the best and worst prizes in $X \backslash D_1$. By Lemma 6, note that $\bar{x} \succeq p$ for all $p \in \Delta X$ and $\bar{y} \succeq p \succeq \underline{y}$ for all $p \in \Delta (X \backslash D_1)$. For any $p \succeq \underline{y}$, define $a_p$ such that $p \sim \bar{x} a_p \underline{y}$ as guaranteed by Lemma 4 (if $\bar{x} \sim \underline{y}$, then set $a_p = 0$ for all $p$). For any $p \preceq \bar{y}$, define $b_p$ such that $p \sim \bar{y} b_p x_*$ (if $\bar{y} \preceq x_*$ or $p \preceq x_*$, then set $b_p = 0$). Now define

$$V(p) := \begin{cases} b_{\underline{y}} + c a_p & \text{for } p \succeq \underline{y} \\ b_p & \text{for } p \preceq \underline{y} \end{cases}$$

where $c = \frac{1 - b_{\underline{y}}}{a_{\bar{y}}}$ if $\bar{y} \succ \underline{y}$ and $c = 1$ otherwise. Note that this is well-defined as for $p \sim \underline{y}$, $a_{\underline{y}} = 0$ so $V(p) = b_{\underline{y}} + c a_{\underline{y}} = b_{\underline{y}}$.

We will show that $V$ represents $\succeq$. First, suppose $p \succeq q$. If $\underline{y} \succeq p$ or $q \succeq \underline{y}$, then $V(p) \geq V(q)$ from Lemma 3. Now, suppose $p \succeq \underline{y} \succeq q$. Note that $b_{\underline{y}} \geq b_q$ so

$$V(p) = b_{\underline{y}} + c a_p \geq b_q + c a_p \geq b_q = V(q)$$

as desired. Finally, suppose $V(p) \geq V(q)$. If $V(q) \geq b_{\underline{y}}$, then $a_p \geq a_q$ so $p \succeq q$. If $b_{\underline{y}} \geq V(p)$, then $b_p \geq b_q$ so again $p \succeq q$. If $V(p) \geq b_{\underline{y}} \geq V(q)$, then $p \succeq \underline{y} \succeq q$ so $p \succeq q$ as desired.

By Axioms 1, 3, and 2, $\succeq$ on $\Delta D_i$ is represented by some

$$\sum_{x \in D_i} q(x) v_i(x).$$

Since $V$ also represents $\succeq$ on $\Delta D_i$, we have for any $q \in \Delta D_i$,

$$V(q) = \psi_i \left( \sum_{x \in D_i} q(x) v_i(x) \right)$$

for some strictly increasing $\psi_i : \mathbb{R} \to \mathbb{R}$. For any $p \in \Delta X$, we can rewrite it as

$$p = \sum_i p_i q_i$$

where $p_i = p(D_i)$ and $q_i \in \Delta D_i$.

First consider the case where $q_1 \succeq \underline{y}$ so $q_i \succeq \underline{y}$ for all $q_i$ by Lemma 6. Thus, $q_i \sim \bar{x} a_{q_i} \underline{y}$ for all $i$. By Lemma 5, we have

$$p = \sum_i p_i q_i \sim \sum_i p_i \left( \bar{x} a_{q_i} \underline{y} \right) = \bar{x} \left( \sum_i p_i a_{q_i} \right) \underline{y}.$$

Since $p \succeq \underline{y}$,

$$\begin{aligned} V(p) = b_{\underline{y}} + c a_p &= b_{\underline{y}} + c \left( \sum_i p_i a_{q_i} \right) \\ &= \sum_i p_i \left( b_{\underline{y}} + c a_{q_i} \right) = \sum_i p_i V(q_i) \\ &= \sum_i p_i \psi_i \left( \sum_{x \in D_i} q_i(x) v_i(x) \right). \end{aligned}$$

Now, consider the case where $q_1 \preceq \underline{y}$ so $\bar{y} \succeq q_i \succeq x_*$ for all $q_i$. Thus, $q_i \sim \bar{y} b_{q_i} x_*$ for all $i$. By Lemma 5 again, we have

$$p = \sum_i p_i q_i \sim \sum_i p_i \left( \bar{y} b_{q_i} x_* \right) = \bar{y} \left( \sum_i p_i b_{q_i} \right) x_*$$

so

$$b_p = \sum_i p_i b_{q_i}.$$

We will show that $V(p) = b_p$. If $p \preceq \underline{y}$, then $V(p) = b_p$ by definition. On the other hand, if $p \succ \underline{y}$, then let $q := \bar{x} a_{\bar{y}} \underline{y} \sim \bar{y}$ and $r := \bar{y} b_{\underline{y}} x_* \sim \underline{y}$ (note that we can assume $\bar{y} \succ \underline{y}$ as if $\bar{y} \sim \underline{y}$, then $p \succ \underline{y}$ cannot arise as $q_1 \preceq \underline{y} \sim \bar{y}$). Now,

$$p \sim \bar{y} b_p x_* = \bar{y} \left( \frac{b_p - b_{\underline{y}}}{1 - b_{\underline{y}}} \right) r \sim q \left( \frac{b_p - b_{\underline{y}}}{1 - b_{\underline{y}}} \right) \underline{y} = \bar{x} \left( \frac{a_{\bar{y}}}{1 - b_{\underline{y}}} \left( b_p - b_{\underline{y}} \right) \right) \underline{y}$$

where the second indifference follows from Lemma 7. Therefore,

$$a_p = \frac{a_{\bar{y}}}{1 - b_{\underline{y}}} \left( b_p - b_{\underline{y}} \right)$$

so

$$V(p) = b_{\underline{y}} + c a_p = b_p$$

as desired. Since $q_i \sim \bar{y} b_{q_i} x_*$ for all $i$, this also implies that $V(q_i) = b_{q_i}$. Thus,

$$V(p) = b_p = \sum_i p_i b_{q_i} = \sum_i p_i V(q_i) = \sum_i p_i \psi_i \left( \sum_{x \in D_i} q_i(x) v_i(x) \right). \tag{2}$$

Next, we show that $\psi_i$ is continuous. Suppose otherwise, so there exists some $t \in \mathbb{R}$ such that

$$V(\bar{x}_i) = \psi_i(v_i(\bar{x}_i)) > t > \psi_i(v_i(\underline{x}_i)) = V(\underline{x}_i)$$

where $\bar{x}_i$ and $\underline{x}_i$ are the best and worst prizes in $D_i$ but $t \neq \psi_i(q_i \cdot v_i)$ for all $q_i \in \Delta D_i$. Suppose there exists a $z \in X \backslash D_i$ such that $V(z) \geq t$. Given (2), we can find some $a \in [0,1]$ such that $V(za\underline{x}_i) = t$ so $\bar{x}_i \succ za\underline{x}_i \succ \underline{x}_i$. But that means there exists a $\lambda \in (0,1)$ such that

$$V(\bar{x}_i \lambda \underline{x}_i) = V(za\underline{x}_i) = t$$

yielding a contradiction. The case for $V(z) < t$ is symmetric.

Finally, define $u(x) := V(\delta_x) = \psi_i(v_i(x))$ for every $x \in D_i$, and let $\phi_i := \psi_i^{-1}$ so

$$V(p) = \sum_i p_i \phi_i^{-1} \left( \sum_{x \in D_i} q_i(x) \phi_i(u(x)) \right),$$

Note the representation is regular by Proposition 4. This concludes the sufficiency proof.

For necessity, note that Proposition 4 implies that $\mathcal{D} = \mathcal{D}_{\succeq}$ must be the partition generated from $\succeq$. Given this, Axioms 1, 3, 4, and 5 are straightforward. To see

Axiom 2, note that

$$
\begin{aligned}
V(paq) &= \sum_i \left( \sum_{y \in D_i} (paq)(y) \right) \phi_i^{-1} \left( \sum_{x \in D_i} \frac{(paq)(x)}{\sum_{y \in D_i} (paq)(y)} \phi_i(u(x)) \right) \\
&= \sum_{i, y \in D_i} (ap(y) + (1-a) q(y)) \phi_i^{-1} \left( \frac{\sum_{x \in D_i} (ap(x) + (1-a) q(x)) \phi_i(u(x))}{\sum_{y \in D_i} (ap(y) + (1-a) q(y))} \right)
\end{aligned}
$$

which is continuous in $a$.

# D Example 4

We verify that $\succeq$ represented by $V$ in Example 4 satisfies Axioms 1-4 but not Axiom 5. First, note that the two formulas defining $V$ coincide when $a \in \{0,1\}$ and when $t_1 = t_2$. Hence, $V$ is well-defined and continuous, so $\succeq$ satisfies Axioms 1 and 2.

We now identify the induced domains. We show that $x_1 \simeq y_1$. First, consider $x_2$ and note that for all $p \in \Delta\{x_1, y_1, x_2\}$,

$$
V(p)^2 = at_1^4 + (1-a)
$$

as $t_1 \leq t_2 = 1$. This is a pure risk representation on $\Delta\{x_1, y_1, x_2\}$ so for any $p, q \in \Delta\{x_1, y_1, x_2\}$, $p \sim q$ implies $p\lambda x_2 \sim q\lambda x_2$. Second, consider $y_2$ and note that for all $p \in \Delta\{x_1, y_1, y_2\}$,

$$
V(p) = at_1^2
$$

as $t_1 \geq t_2 = 0$. This is a pure risk representation on $\Delta\{x_1, y_1, y_2\}$ so for any $p, q \in \Delta\{x_1, y_1, y_2\}$, $p \sim q$ implies $p\lambda y_2 \sim q\lambda y_2$. This establishes that $x_1 \simeq y_1$. The case for showing $x_2 \simeq y_2$ is symmetric.

We now show that no cross pair is related. Note that $x_1 \sim x_2$ and $y_1 \sim y_2$, but

$$
\begin{aligned}
V\left(x_1 \frac{1}{2} y_2\right) &= \frac{1}{2} \neq \left(\frac{1}{2}\right)^2 = V\left(x_2 \frac{1}{2} y_2\right) \\
V\left(y_1 \frac{1}{2} x_2\right) &= \sqrt{\frac{1}{2}} \neq \left(\frac{1}{2}\right)^2 = V\left(y_2 \frac{1}{2} x_2\right)
\end{aligned}
$$

which establishes that $x_1 \not\simeq x_2$ and $y_1 \not\simeq y_2$. Next, let $q = x_2 \frac{1}{2} y_2$ and note that

$p = x_1 \frac{1}{4} y_2 \sim q$ and $r = y_1 \frac{15}{16} x_2 \sim q$. However,

$$V\left(p\frac{2}{3}x_2\right) = \tfrac{3}{10} \neq \tfrac{4}{9} = V\left(q\frac{2}{3}x_2\right)$$
$$V\left(r\frac{4}{5}y_2\right) = \tfrac{1}{50} \neq \tfrac{4}{25} = V\left(q\frac{4}{5}y_2\right)$$

which establishes that $x_1 \not\sim y_2$ and $y_1 \not\sim x_2$. Taken together, this implies

$$\mathcal{D}_{\succeq} = \{D_1, D_2\}.$$

It is straightforward to see that Axiom 3 is satisfied. For Axiom 4, let $(p, a, r)$ and $(q, a, r)$ be pure inter-domain mixtures, and suppose $p \succeq q$. Since $(p, a, r)$ is a pure inter-domain mixture, Definition 5 requires $p$, $r$, and $par$ to have the same conditional lotteries (if defined) on $D_1$ and $D_2$. By the same argument, $q$, $r$, and $qar$ have the same conditional lotteries (if defined) on $D_1$ and $D_2$. First, note that if $r(D_i) > 0$ for both $i \in \{1, 2\}$, then $p$ and $q$ must have the same conditional lotteries (if defined) as $r$. Thus, either $t_1 \geq t_2$ or $t_1 \leq t_2$ for all lotteries $p$, $q$, $r$, $par$ and $qar$. In the case of the former, we have

$$V(par) = aV(p) + (1-a)V(r)$$
$$V(qar) = aV(q) + (1-a)V(r)$$

so the result follows. In the case of the latter, we have

$$V(par)^2 = aV(p)^2 + (1-a)V(r)^2$$
$$V(qar)^2 = aV(q)^2 + (1-a)V(r)^2$$

so again the result follows. Now, suppose $r(D_2) = 1$ and note that $V(r) = r(x_2)^2$. If $V(p) \geq V(q) \geq r(x_2)^2$, then we have the same set of equations as the former case above so the result follows. If $r(x_2)^2 \geq V(p) \geq V(q)$, then we have the same equations as the latter case above so again the result follows. Finally, suppose $V(p) > r(x_2)^2 > V(q)$. In this case, since $V(r) = r(x_2)^2$,

$$V(par) = aV(p) + (1-a)V(r) > V(r) > \sqrt{aV(q)^2 + (1-a)V(r)^2} = V(qar)$$

as desired. The case for $r(D_1) = 1$ is symmetric. This means that Axiom 4 holds.

Finally, $\delta_{x_1} \sim \delta_{x_2}$ and $\delta_{y_1} \sim \delta_{y_2}$, whereas

$$V\left(\frac{1}{2}\delta_{x_1} + \frac{1}{2}\delta_{y_2}\right) = \frac{1}{2} \neq \sqrt{\frac{1}{2}} = V\left(\frac{1}{2}\delta_{x_2} + \frac{1}{2}\delta_{y_1}\right).$$

Hence Axiom 5 is violated, while Axioms 1-4 are satisfied.